\documentclass[12pt,a4paper,final]{iopart}

\usepackage{iopams}  
\usepackage{graphicx}

\usepackage{subcaption} 

\usepackage[breaklinks=true,colorlinks=true,linkcolor=blue,urlcolor=blue,citecolor=blue]{hyperref}

\usepackage{cancel}

\newcommand{\bea}{\begin{eqnarray}}
\newcommand{\ea}{\end{eqnarray}}
\newcommand{\eea}{\end{eqnarray}}

\begin{document}

\title[Martingale Strategy for Modeling Quantum Adiabatic Evolution]{Martingale Strategy for Modeling Quantum Adiabatic Evolution}

\author{Elnaz Darsheshdar}
\address{Institute of Physics, Polish Academy of Sciences,
  Aleja Lotnikow 32/46, PL-02668 Warsaw, Poland}
\ead{elnaz@ifpan.edu.pl, Darsheshdare@gmail.com}

\author{Seyed Mostafa Moniri}
\address{Department of Engineering Science, Golpayegan University of Technology, Golpayegan, Iran}
\ead{moniri@gut.ac.ir}

\author{Patrick Navez}
\address{
Skolkovo Institute of Science and Technology, 
Skolkovo Innovation Centre, Nobel Street 3, Moscow 143026, Russia\\
University of Saskatchewan, Dept of Math. and Stat. , Saskatoon, S7N 5E6, Canada}
\ead{navez@physics.uoc.gr}

\author[cor1]{Alexandre Zagoskin}
\address{Department of Physics, Loughborough University, Loughborough LE11 3TU, United Kingdom\\
National University of Science and Technology "MISIS", Leninsky prosp. 4, 
119049 Moscow, Russia}
\ead{a.zagoskin@lboro.ac.uk}

\begin{abstract}
We propose a strategy for modeling the behaviour of an adiabatic quantum computer described by an Ising Hamiltonian with $N$ sites and the coordination number $Z$. The method is based on the $1/Z$-expansion for the density matrix of the system. In each order, the ground state energy is found neglecting the higher-order correlations between the sites, as long as the set of equations remains non-singular. The conditions of the appearance of a singularity, equivalent to the disappearance of energy gap in the given approximation, can be directly obtained from the equations. Then the next order in the expansion must be used, at the price of an $N$-fold increase in computational resources. This "martingale" strategy allows reducing the computational costs to a power of $N$ rather than $2^N$, with a finite probability of success. 
The strategy is illustrated by the case of a two-spin system and extended to a large number of qubits. Comparing the predictions to the experimental results obtained by using an adiabatic quantum computer would help quantify the importance of multi-site correlations, and the influence of decoherence, on its operation. 
\end{abstract}

\vspace{2pc}

\section{Introduction}

The enormous progress in classical computation (both hardware and algorithms) still leaves the large-scale NP problems (e.g., factorizing, breaking RSA encryption, travelling salesman) intractable due to the exponential growth of the computational costs with the input size. The existence of this barrier (insurmountable in practice, and if the conjecture $NP \neq P$ is proven, in principle)  requires either a different strategy of classical computations, or a fundamentally new technology such as a quantum computer \cite{Nielsen}.  The ability of a universal quantum computer to perform in the presence of decoherence, the scale at which it would outperform a classical computer, and a way to determine whether such an outperformance took place and to what extent, remain the area of controversy and very active research   \cite{Ball2017,Neill2017,Pednault2017}. 

As far as  adiabatic quantum computers are concerned \cite{KN98,Farhi,google,exp2,exp3,exp4}, 
the degree of entanglement and quantum coherence essential for their operation or for the exponential speedup compared to classical computers has not been experimentally established either, and several tests concerning the detection and evaluation of quantum correlations in these systems have been proposed \cite{L14,Navez_2017}. Same as with universal quantum computers, the major obstacle lies in the necessity to simulate the behaviour of a large quantum system with classical means, and accurately enough to make the comparison with an experiment meaningful\cite{Zagos}. 

A direct simulation of an $N$-qubit device would require dealing with a $2^N$-dimensional Hilbert space, which, while allowing to account for all the quantum correlations existing in the system, puts it out of reach for any feasible classical computation for $N \sim 1000$ (which is an overoptimistic assumption). On the other hand, quantum field theoretical treatments of macroscopic quantum systems ($N \sim 10^{23}$), which take into account only few-point correlations, are successful and efficient in describing a wide range of phenomena. 

In this paper we propose a strategy based on using an adaptive algorithm for finding the ground state of a quantum coherent system of $N \gg 1$ qubits undergoing an adiabatic evolution, which takes into account correlations between the smallest possible number of qubits, and increases this number only after the calculations break down. The condition for this breakdown (\textsl{singularity condition}) can be directly obtained from the equations and corresponds to the disappearance of the energy gap between the ground and first excited state of the system in the given order of approximation. The transition procedure to the next order is well defined. Each transition multiplies the computation costs by $N$, but the probability of finding the ground state before running out of resources remains finite. This is similar to the martingale strategy of betting (doubling the stakes after each loss).

The residual success rate  depends on the number of  qubits and can serve as a benchmark test of the performance of a quantum device by determining indirectly the role played by the decoherence processes. Decoherence tends to suppress and   eventually destroy quantum correlations between qubits; thus the comparison between the success rates predicted in different orders of approximation and that observed in an actual adiabatic quantum computer would provide a measure of the amount of quantum correlations present in the system, as well as indicate the minimal degree of these correlations required for the operation of this device.

To be specific, we use the standard description of an adiabatic quantum computer (Ising Hamiltonian with the coordination number $Z$ in an external field) with $Z \gg 1$, and apply the $1/Z$ expansion \cite{Navez_2010,Navez_2014,Navez_2016,Navez_20104}.  This produces a hierarchy of dynamical equations for $n$-site reduced density matrices allowing to 
systematically describe the adiabatic dynamics, taking into account the desired level of correlations. Each higher order  requires an $N$-fold increase of computational resources.  

The paper is organized as follows. In Section 2, the principle of adiabatic quantum computation is briefly outlined. In Section 3,  after a brief introduction to $1/Z$-expansion, the hierarchy of equations for the reduced density matrices is derived, the singularity criterion is established, and indicates the disappearance of the energy gap between the ground and first excited state in given approximation. The efficiency of the approach in the lowest order approach is illustrated in Section 4 for a two-qubit system, and in Section 5 for a multiqubit system. Section 6 contains conclusions and a discussion of the perspectives of the proposed method.

\section{Definitions}

Under quite general assumptions, the operation of an adiabatic quantum computer can be reduced finding  the 
ground state of an $N$-site Ising Hamiltonian\cite{Nielsen,KN98,Farhi,exp3}. For interesting cases the spin-spin couplings are nontrivially distributed, producing spin glass-like behaviour and making finding this minimum by classical means a difficult (NP) problem. Basically, one has to find the state delivering the absolute minimum to the 
energy function
\begin{eqnarray} \label{eq:CL}
E=-\frac{1}{Z}\sum\limits_{\nu, \mu = 1}^{N}{{{J}_{\nu \mu }}{S}_{\mu }^{z}{S}_{\nu }^{z}}-\sum\limits_{\mu =1}^{N}{{{J}_{\mu }}{S}_{\mu }^{z}}
\end{eqnarray} 
where ${S}_{\mu }^{z}=\pm \frac{1}{2}$ is a classical bit at a site $\mu$ 
and ${{J}_{\mu\nu }} = {{J}_{\nu\mu }}$, ${{J}_{\mu }}$ are spin-spin and spin-field couplings respectively. The coordination number $Z$ is the number of nonzero $J_{\mu\nu}$ for any given $\mu$. 

Replacing in (\ref{eq:CL}) classical bits $S_{\mu}^z$, with quantum bits (1/2-spins) $\hat{S}_{\mu}^z$, we obtain a Hamiltonian $\hat{H}_f$ commuting with each $\hat{S}_{\mu}^z$, and therefore not inducing any dynamics. In order to unfreeze the system one can add to $\hat{H}_f$ a non-commuting term, e.g.,  $\hat{H}_B=-B\sum\limits_{\mu }{\hat{S}_{\mu }^{x}}$, to obtain
\begin{eqnarray}
\hat{H}=-\frac{1}{Z}\sum\limits_{\nu \mu \in N}{{{J}_{\nu \mu }}\hat{S}_{\mu }^{z}\hat{S}_{\nu }^{z}}-
\sum\limits_{\mu \in N}{{{J}_{\mu }}\hat{S}_{\mu }^{z}}-B\sum\limits_{\mu }{\hat{S}_{\mu }^{x}}
\end{eqnarray} 
The ground state of $\hat{H}_B$ is obviously a factorized eigenstate of every $\hat{S}^x_{\mu}$ with $\langle\hat{S}^x_{\mu}\rangle = 1/2$ (if $B > 0$).  
If now include in the parameters ${{J}_{\mu\nu }}$, ${{J}_{\mu }}$, $B$ a slow dependence on time such that, e.g., at $\hat{H}(t=-\infty) = \hat{H}_B,$ while $\hat{H}(t=0) = \hat{H}_f$, and initialize the system in the ground state of $\hat{H}_B$, then by the virtue of the adiabatic theorem at $t=0$ the system will be in the (factorized) ground state of $\hat{H}_f$ thus solving the optimization problem. This is the essence of adiabatic quantum computing. (See \cite{Albash2018} for a detailed review and a discussion of the conditions when this approach is applicable.)

\section{Mean field adiabatic equations}

The dynamics of general lattice Hamiltonians is addressed, e.g., in \cite{Navez_2014,QKNS14}. The time evolution 
of the density matrix of the system is governed by the von Neumann-Liouville equation 
$i\hbar {{\partial }_{t}}\hat{\rho} =\left[ \hat{H},\hat{\rho}  \right]$. 

In order to simplify the analysis one routinely introduces the set of reduced density matrices, 
${{\hat{\rho }}_{\mathcal{S}}}=\rm{Tr}_{\cancel{\mathcal{S}}}\hat{\rho }$, which is obtained by tracing out the Hilbert 
spaces of all sites except a few $\mathcal{S}=\left\{ {{\mu }_{1}},\,{{\mu }_{2}},\,...\,,\,{{\mu }_{n}} \right\}$. 
If we keep only one site $\mu$, the reduced density matrix ${{\hat{\rho} }_{\mu }}$ is a linear operator acting 
on the smaller Hilbert space of one spin; if we keep two sites $\mu $, $\nu $, then ${{\hat{\rho} }_{\mu \nu }}$ 
exists in the two-spin Hilbert space etc. The decomposition 
${{\hat{\rho} }_{\mu \nu }}=\hat{\rho} _{\mu \nu }^{corr}+{{\hat{\rho} }_{\mu }}{{\hat{\rho} }_{\nu }}$, and 
${{\hat{\rho} }_{\mu \nu \lambda }}=\hat{\rho} _{\mu \nu \lambda }^{corr}+\hat{\rho} _{\mu \nu }^{corr}{{\hat{\rho} }_{\lambda }}+
\hat{\rho} _{\mu \lambda }^{corr}{{\hat{\rho} }_{\nu }}+\hat{\rho} _{\lambda \nu }^{corr}{{\hat{\rho} }_{\mu }}+
{{\hat{\rho} }_{\mu }}{{\hat{\rho} }_{\nu }}{{\hat{\rho} }_{\lambda }}$ etc. allows to derive an exact hierarchy of 
interlinked equations for these operators equivalent to the original Liouville - von Neumann equation. It has the advantage of directly producing approximations with any desired degree of multispin correlations.

The large coordination number expansion assumes that the coordination number $Z\gg 1$, 
 so that higher order terms give decreasingly smaller contributions to the lattice system dynamics, 
  ${{\hat{\rho }}_{\mathcal{S} \cup {\mu }_{n+1} }} \sim {{\hat{\rho }}_{\mathcal{S}}}/Z.$
 The hierarchy of reduced density matrices allow us to systematically determine the equilibrium 
properties such as the ground state \cite {Navez_2017, Navez_2016} as well as non-equilibrium dynamics \cite {Navez_2010, Navez_20104}. In particular, it provides a method of  
finding the ground state and the 
quench dynamics of a uniform quantum Ising model in any  dimension, including the quantum phase transition 
between the paramagnetic and ferromagnetic phases and the excitation energy spectrum, 
and the quench dynamics of a uniform quantum Ising model \cite{Navez_2017}.

The equations up to first order have the following form:
\begin{eqnarray}\label{eq:leading}
i{{\partial }_{t}}{{\hat{\rho} }_{\mu }}=\frac{1}{Z}\sum\limits_{\kappa \ne \mu }{\rm{Tr}_{\kappa }
\left\{ \hat{\mathcal{L}}_{\mu \kappa }^{S}\left( \hat{\rho} _{\mu \kappa }^{corr}+{{\hat{\rho} }_{\mu }}
{{\hat{\rho} }_{\kappa }} \right) \right\}}+{{\hat{\mathcal{L}}}_{\mu }}{{\hat{\rho} }_{\mu }}
\end{eqnarray}
\begin{eqnarray}
&& i{{\partial}_{t}}\hat{\rho} _{\mu \nu }^{corr}={{\hat{\mathcal{L}}}_{\mu }}\hat{\rho} _{\mu \nu }^{corr}+
\frac{1}{Z}{{\hat{\mathcal{L}}}_{\mu \nu }}{{\hat{\rho} }_{\mu }}{{\hat{\rho} }_{\nu }}-\frac{{{\hat{\rho} }_{\mu }}}{Z}\rm{Tr}_{\mu }\left\{ \hat{\mathcal{L}}_{\mu \nu }^{S}{{\hat{\rho} }_{\mu }}{{\hat{\rho} }_{\nu }} \right\} \\ \nonumber
&& +\frac{1}{Z}\sum\limits_{\kappa \ne \mu ,\nu }{\rm{Tr}_{\kappa }\left\{ \hat{\mathcal{L}}_{\mu \kappa }^{S}\left( \hat{\rho} _{\mu \nu }^{corr}{{\hat{\rho} }_{\kappa }}+\hat{\rho} _{\nu \kappa }^{corr}{{\hat{\rho} }_{\mu }} \right) \right\}} +\left( \mu \leftrightarrow \nu  \right)+\mathcal{O}(1/{{Z}^{2}}) 
\end{eqnarray} 
while we have $\hat{\mathcal{L}}_{\mu \nu }^{S}=\hat{\mathcal{L}}{{}_{\mu \nu }}+{{\hat{\mathcal{L}}}_{\nu \mu }}$ and 
the Liouville operators are ${{\hat{\mathcal{L}}}_{\mu }}\hat{\rho} =[-{{{J}_{\mu }}\hat{S}_{\mu }^{z}}-B{\hat{S}_{\mu }^{x}} ,\hat{\rho} ]$ 
and ${{\hat{\mathcal{L}}}_{\mu \nu }}\hat{\rho} =[-{{{J}_{\nu \mu }}\hat{S}_{\mu }^{z}\hat{S}_{\nu }^{z}},\hat{\rho} ]$.

In this work, we shall initially use the leading order  to determine the $z$-components of lattice spins at the end of the adiabatic operation.  Neglecting two particle correlation in (\ref{eq:leading}) produces  closed equations. We define the spin expectation value as 
$S_{\mu }^{i}=\left\langle \hat{S}_{\mu }^{i} \right\rangle ={\rm Tr}(\hat{S}_{\mu }^{i}\hat{\rho} )={\rm Tr}_\mu (\hat{S}_{\mu }^{i}\hat{\rho}_\mu )$. 
Taking  these expectations in (\ref{eq:leading}), we obtain after straightforward algebraic calculations the system of equations: 
\begin{eqnarray} \label{eq:mf11}
\partial_t S_\mu^x&=&\frac{2}{Z}\sum_{\nu \not=\mu} J_{\mu \nu} (t)S_\mu^y  S_\nu^z + J_\mu (t) S_\mu^y
\\
\partial_t S_\mu^y&=&-\frac{2}{Z}\sum_{\nu \not=\mu} J_{\mu \nu} (t) S_\mu^x S_\nu^z + B(t) S_\mu^z -  J_\mu (t) S_\mu^x
\\
\partial_t S_\mu^z&=&- B(t) S_\mu^y 
\end{eqnarray} 
The  polarization of the transverse magnetic field imposes the initial condition  
${S}_{\mu}^x=\frac{1}{2}$, ${S}_{\mu}^y=S_{\mu}^z=0$. In order to emulate adiabatic switching between the initial and final Hamiltonians, we  have parametrized the coupling and external field terms via ${{J}_{\mu \nu }}\to J_{\mu\nu}(t) \equiv s(t){{J}_{\mu \nu }}$ , ${{J}_{\nu }}\to J_{\nu}(t) \equiv s(t){{J}_{\nu }}$, $B \to B(t) \equiv 1-s(t)$. The explicit dependence $s(t)$ is chosen, like in \cite{Navez_2017}:   $$s(t)=\exp(\epsilon t); \\ t=]-\infty, 0]; \\  \epsilon\to 0.$$
The adiabatic evolution is monitored in the range  $s \in [0,1]$. 

Since $\partial_t =\epsilon s\partial_s$, the time parameter 
is eliminated from the dynamical equations:
\begin{eqnarray} \label{eq:mf1}
&&\epsilon s \partial_s S_\mu^x=\frac{2s}{Z}\sum_{\nu \not=\mu} J_{\mu \nu} S_\mu^y  S_\nu^z + sJ_\mu S_\mu^y
\\
\label{eq:syzp}
&&\epsilon s \partial_s S_\mu^y=-\frac{2s}{Z}\sum_{\nu \not=\mu} J_{\mu \nu} S_\mu^x S_\nu^z + (1-s) S_\mu^z -  sJ_\mu S_\mu^x
\\
&&\epsilon s \partial_s  S_\mu^z=- (1-s) S_\mu^y 
\end{eqnarray}
In the adiabatic limit ($\epsilon \rightarrow 0$),
the choice of the scaling $S^{y}_{\mu} =O{(\epsilon)}$ and the scaling unity   
for all other dynamical variables allows to obtain $\epsilon$-independent equation. Indeed the elimination of $S^{x}_{\mu}$ and 
$S^{y}_{\mu}$ using (\ref{eq:mf1}) and  (\ref{eq:syzp}) together with the limit $\epsilon \rightarrow 0$ results in:
\begin{eqnarray}\label{eq:mf3p}
(1+A_\mu^2)A_\mu \partial_s S_\mu^z=S_\mu^z \frac{ d A_\mu}{ds}
\end{eqnarray}
with $A_\mu = [\frac{2}{Z}\sum_{\nu \not=\mu}  J_{\mu \nu} S_\nu^z + J_\mu ]s/(1-s)$. 
We deduce also  
$S^x_\mu = S^z_\mu/A_\mu$ which imposes the initial condition $S_\mu^z\stackrel{s\rightarrow 0}{=}sJ_\mu/2$.
Using these last expressions for the spin components, the total mean field energy is determined from: 
\begin{eqnarray} \label{eq:2}
{{E}_{MF}(s)}=-(1-s)\sum\nolimits_{\mu }{S_{\mu }^{x}}-\frac{s}{Z}\sum\nolimits_{\mu \nu }{{{J}_{\nu \mu }}S_{\mu }^{z}S_{\nu }^{z}}-
s\sum\nolimits_{\mu }{{{J}_{\mu }}S_{\mu }^{z}} 
\end{eqnarray}
and allows to find the ground state energy $E_0=E_{MF}(s=1)$ in the mean field approximation. Note that the expression for $A_\mu$ contains a factor which diverges as $s\to 1$. In general, one should check whether the system (\ref{eq:mf3p}) does not become singular for some $s \in ]0,1[$: otherwise the lowest-order solution (\ref{eq:2}) is wrong. 

In order to analyze these singularities, the Eq.(\ref{eq:mf3p}) can be rewritten  using a more explicit form:
\begin{eqnarray}\label{eq:sing}
\sum_\nu M_{\mu\nu} \partial_s S_\nu^z=
S_\mu^z \left. \frac{\partial A_\mu}{\partial s}\right|_{S_\nu^z}
\end{eqnarray}
where 
\begin{eqnarray}
{{M}_{\mu \nu }}=(1+\frac{4S{{_{\mu }^{z}}^{2}}}{1-\,4S{{_{\mu }^{z}}^{2}}})\frac{2S_{\mu }^{z}}
{\sqrt{1-\,4S{{_{\mu }^{z}}^{2}}}}{{\delta }_{\mu \nu }} -S_{\mu }^{z}\frac{2s}{Z}\sum\limits_{\nu \not{=}\mu }{{{J}_{\mu \nu }}}/(1-s) 
\end{eqnarray}
from which we deduce {\em the singularity condition} 
\begin{equation}
{\rm det}(M_{\mu \nu})=0.
\label{eq:singularity-condition}
\end{equation}

In the nonsingular case  the Eq.(\ref{eq:mf3p}) is integrated exactly resulting in an implicit equation for the spin $z$-components:
\begin{eqnarray} \label{eq:mf4}
S_\mu^z= \frac{A_\mu}{2\sqrt{1+A_\mu^2}}
\end{eqnarray}
In this case we obtain at the end of adiabatic operation ($s=1$):
\begin{eqnarray} \label{eq:spin}
S_{\mu }^{z}=\frac{\frac{2}{Z}\sum\nolimits_{\mu \ne \nu }{{{J}_{\mu \nu }}S_{\nu }^{z}+{{J}_{\mu }}}}
{2\left| \frac{2}{Z}\sum\nolimits_{\mu \ne \nu }{{{J}_{\mu \nu }}S_{\nu }^{z}+{{J}_{\mu }}} \right|}
\end{eqnarray}
This last equation is implicit and cannot be solved for a large 
size systems because it has the  exponential number  $2^N$ of trials $S_\mu^z=\pm 1/2$. Therefore,
instead of going back to an exponential size problem, we will
integrate Eq.(\ref{eq:mf3p}) numerically. This approach is more convenient for detecting the singularities as well.

The singularity condition (\ref{eq:singularity-condition}) is identical to the {\it gapless} condition  that imposes a zero gap in the excitation spectrum. 
Indeed, let us assume a small perturbation of spin around the ground state spin $S_\mu^i$ in Eqs.(\ref{eq:mf11}) of the form 
$S_\mu^i+\delta S_\mu^i e^{i\omega t}$. 
After linearization around the steady ground state solution, we find linear equations for the perturbation $\delta S_\mu^i$. 

The gapless limit condition $\omega \rightarrow 0$ 
imposes $\delta S^{y}_{\mu}=0$. Knowing that  $S^{y}_{\mu}=0$ for the ground state 
and using the normalization condition $S_\mu^x \delta S_\mu^x + \delta S_\mu^z S_\mu^z=0$, we find the matrix equation form 
$\sum_\nu M_{\mu\nu} \delta S_\nu^z=0$ which possess a non trivial solution only if the singularity 
condition is fulfilled. Thus, a singularity in the leading order equations in the $1/Z$ expansion 
is caused by a zero gap in the energy spectrum  whatever  the number of qu-bits involved. As a results, 
the optimization problem cannot be addressed since the mean field equations cannot be resolved unambiguously beyond the singular points $s$.  

In this case, one should consider next order in the $1/Z$ expansion method, hoping for a better chance of success. 
but at the price of $N$-fold increase in the computation time. 
This strategy however diminishes the cost of computation from an exponential cost of ${{2}^{N}}$ to a polynomial one with a decreasing, but finite, probability of success before the computation costs exceed the available resources, like in the  martingale strategy in games of hazard. On the other hand,
this approach  serves also as a basis for  testing any quantum coherent structure used for  adiabatic quantum computation. 
A comparison of the success rate between the theory and the experiment  allows to assess the performance of these devices and, in particular, to determine how it depends on the existence and robustness of multiqubit correlations.

\section{Two-Spin System}

\begin{figure} 
\centering
\includegraphics[scale=.5]{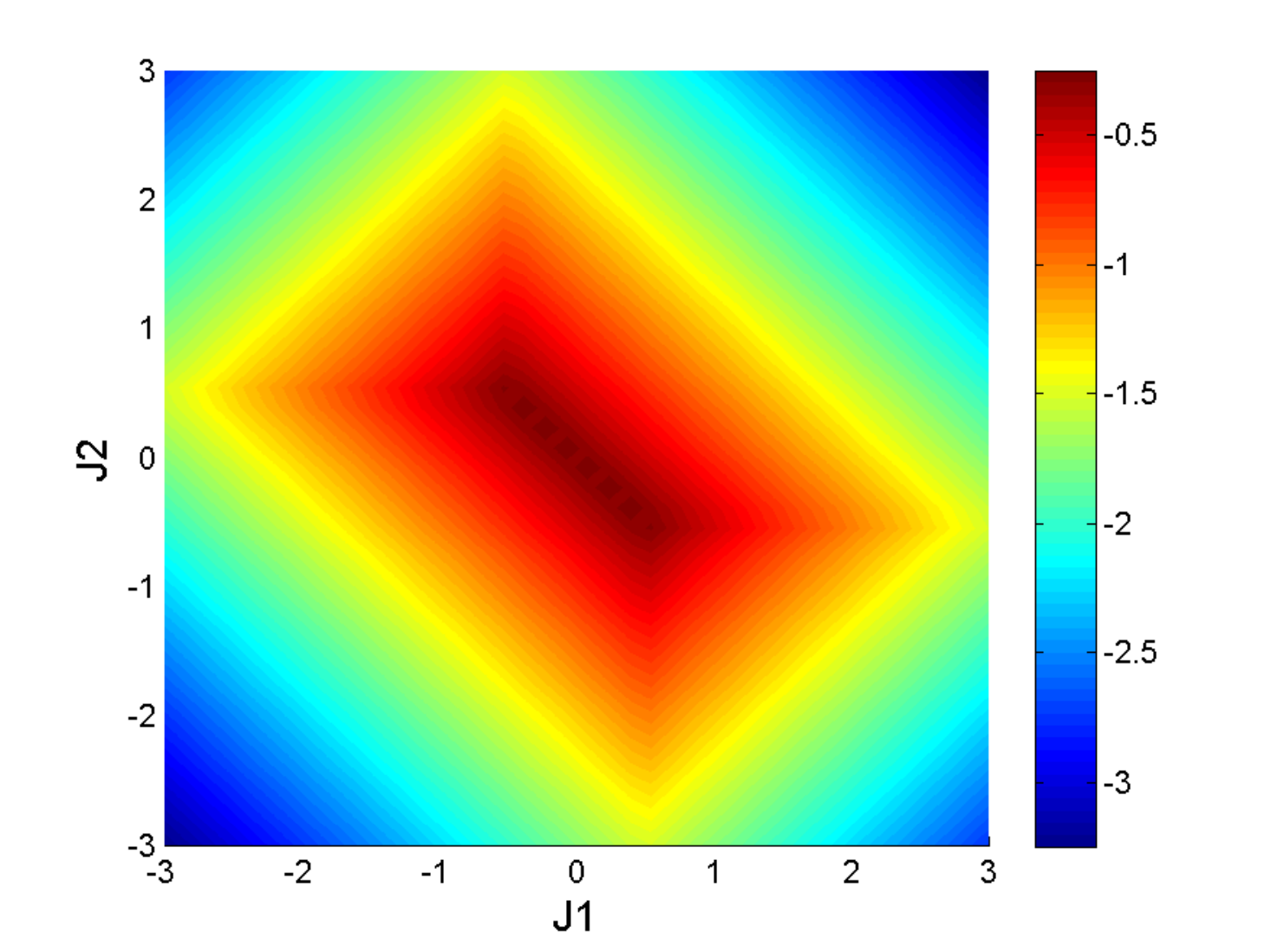}
\caption{Minimum energy $E_0$ of classical Ising Hamiltonian in two spins system.}
\label{fig:2}
\end{figure}

\begin{figure} 
\centering
\includegraphics [scale=.5] {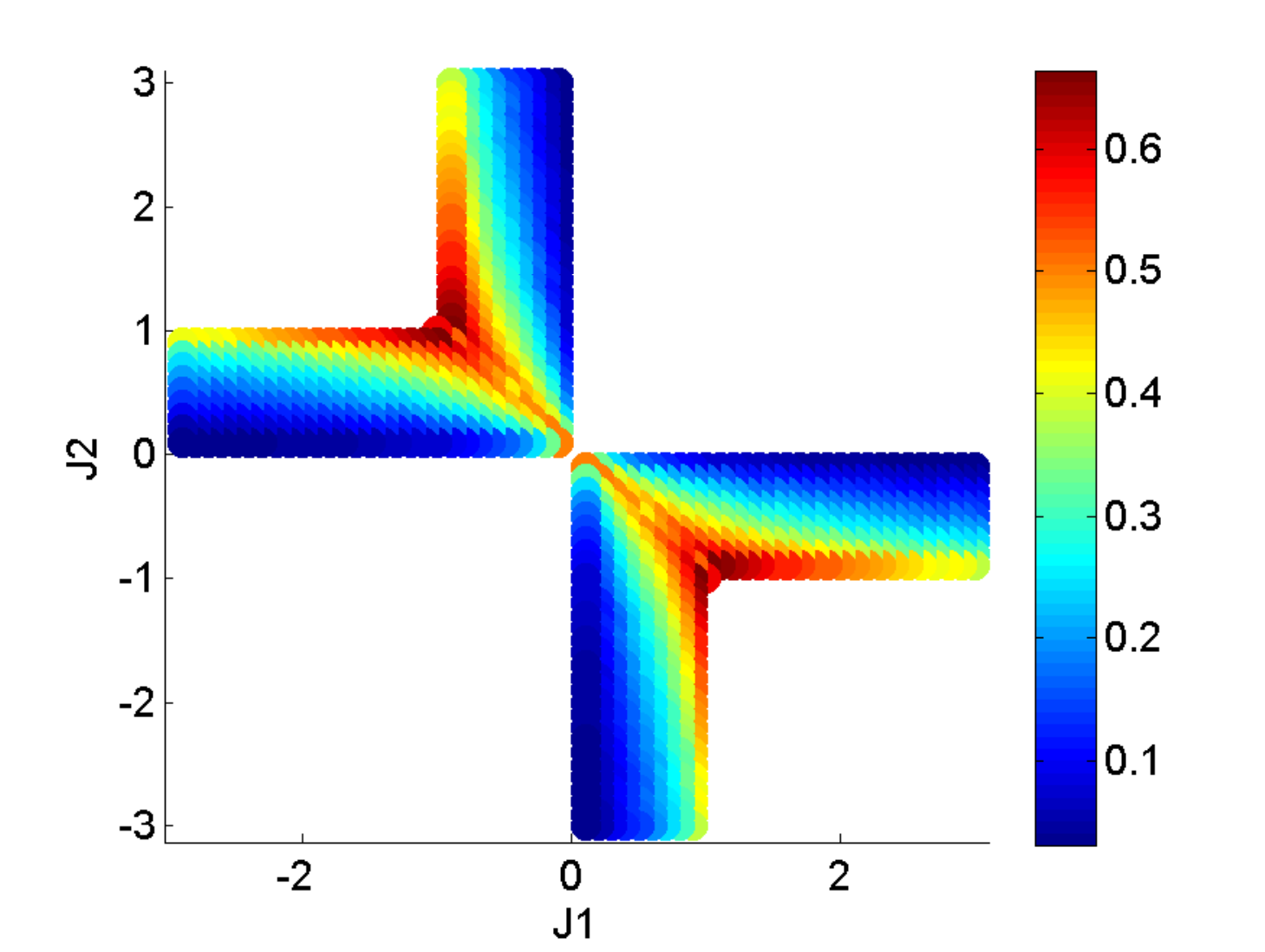}

\caption{The singularity values $s$ within the range of ${{J}_{1}},{{J}_{2}}$ for which it appears.}
\label{fig:region}
\end{figure}

The simple case of just two spins, $\mu=1,2,$  provides instructive insights into the performance of the proposed approach, even though it is rigorously speaking non-applicable ($Z=1$).
We set the values $J_{12}=J_{21}=1$ and change only  ${{J}_{i}}$. The optimized energy for the
classical two-spin Ising Hamiltonian   is shown in Fig. \ref{fig:2} and already displays a nontrivial structure. 
In Fig. \ref{fig:region}, we plot  the values of $s$ inside the region where a singularity occurs. 
In the specified values range of on-site interactions ($J_i=-3 \dots 3$) the lowest order mean field approach succeeds in solving the optimization problem in $26/36=72,2 \%$ of the cases.

\begin{figure} 
\centering
\includegraphics[scale=.6]{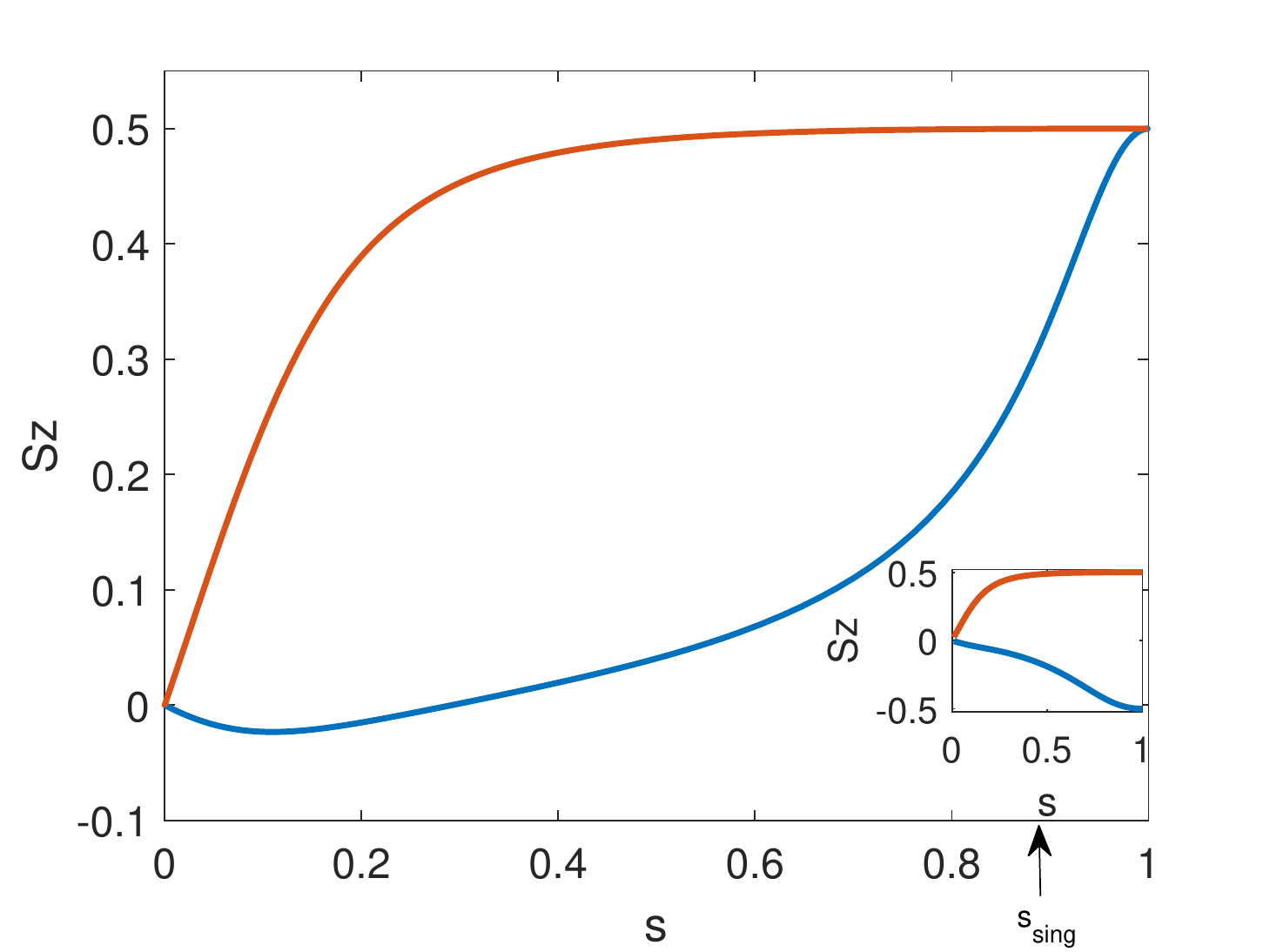}
\caption{Adiabatic evolution in the singular case for ${{J}_{1}}=-0.9,{{J}_{2}}=3,{{J}_{12}}=1$, $S_{1}^{z}\,$ in the blue curve and $S_{2}^{z}\,$ 
in the  red curve. Singularity happens on $s_{sing}=0.89$ . The inset shows results obtained 
from the exact diagonalization where, in contrast, the spins do not change its sign.}
\label{fig:10}
\end{figure}

The mean field  dynamics is compared with the one obtained from exact diagonalization. The 
Fig. \ref{fig:10} shows the mean-field dynamics of the spins during the adiabatic operation in a singular case, according to which one spin changes sign. On the contrary, the exact solution (inset) does not have any spin sign changes during the adiabatic operation. The failure of the lowest order  mean field approximation  is to be expected at a singularity.



 

\begin{figure}
\centering
\begin{subfigure}{.5\textwidth}
  \centering
  \includegraphics[scale=.5]{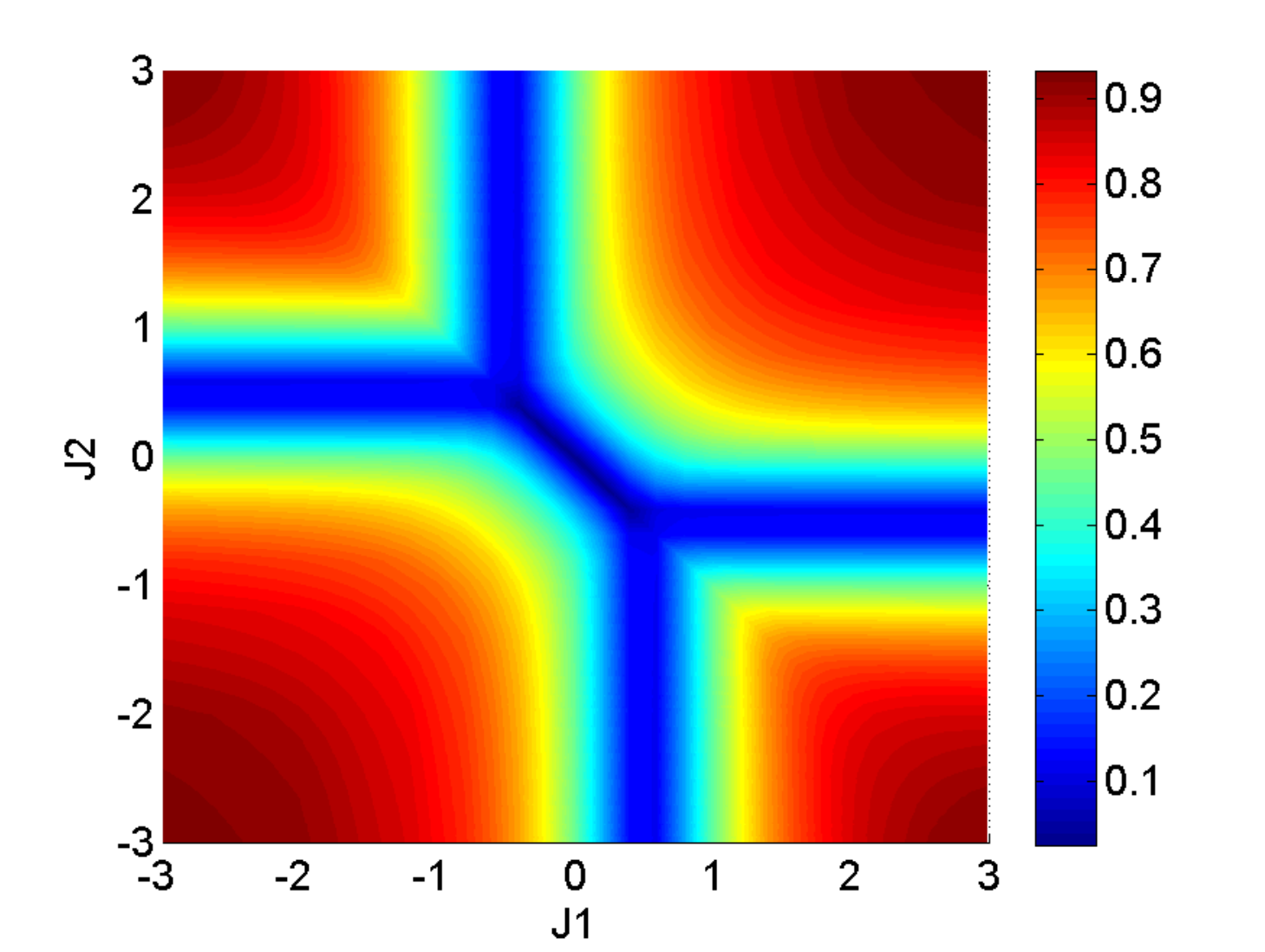}
  \caption{}
  \label{fig:202}
\end{subfigure}%
\begin{subfigure}{.5\textwidth}
  \centering
  \includegraphics[scale=.5]{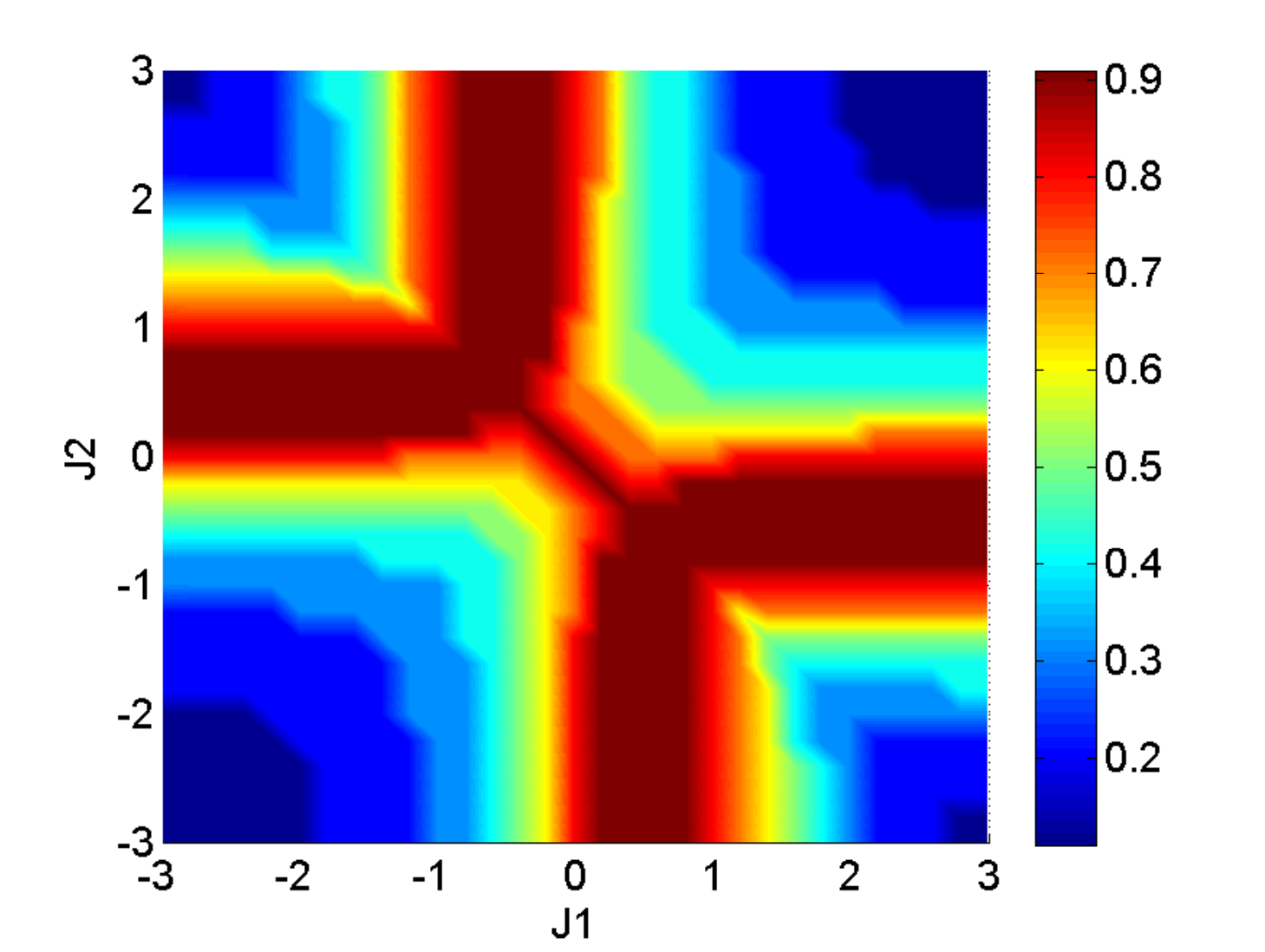}
  \caption{}
  \label{fig:203}
\end{subfigure}
\caption{(a) Minimum energy gap and (b) ${{s}_{gap}}$, as a function of ${{J}_{1}}$ and ${{J}_{2}}$  (exact diagonalization).}
\label{fig:202-203}
\end{figure}

For a general comparison with Figs.\ref{fig:2} and \ref{fig:region}, 
we plot the minimum energy gap in Fig.\ref{fig:202} and the  corresponding value of ${{s}_{gap}}$ at this minimum energy gap in Fig.\ref{fig:203} 
obtained from the exact diagonalization of Hamiltonian. 
The minimum energy gap is always nonzero so that the adiabatic process always solves  the optimization.
Note that smaller gaps happen for higher ${{s}_{gap}}$ values and occurs preferentially in the regions where the mean field approach presents a singularity.

\section{Multiqubit Case}

\begin{figure} 
\centering
\includegraphics[scale=.5]{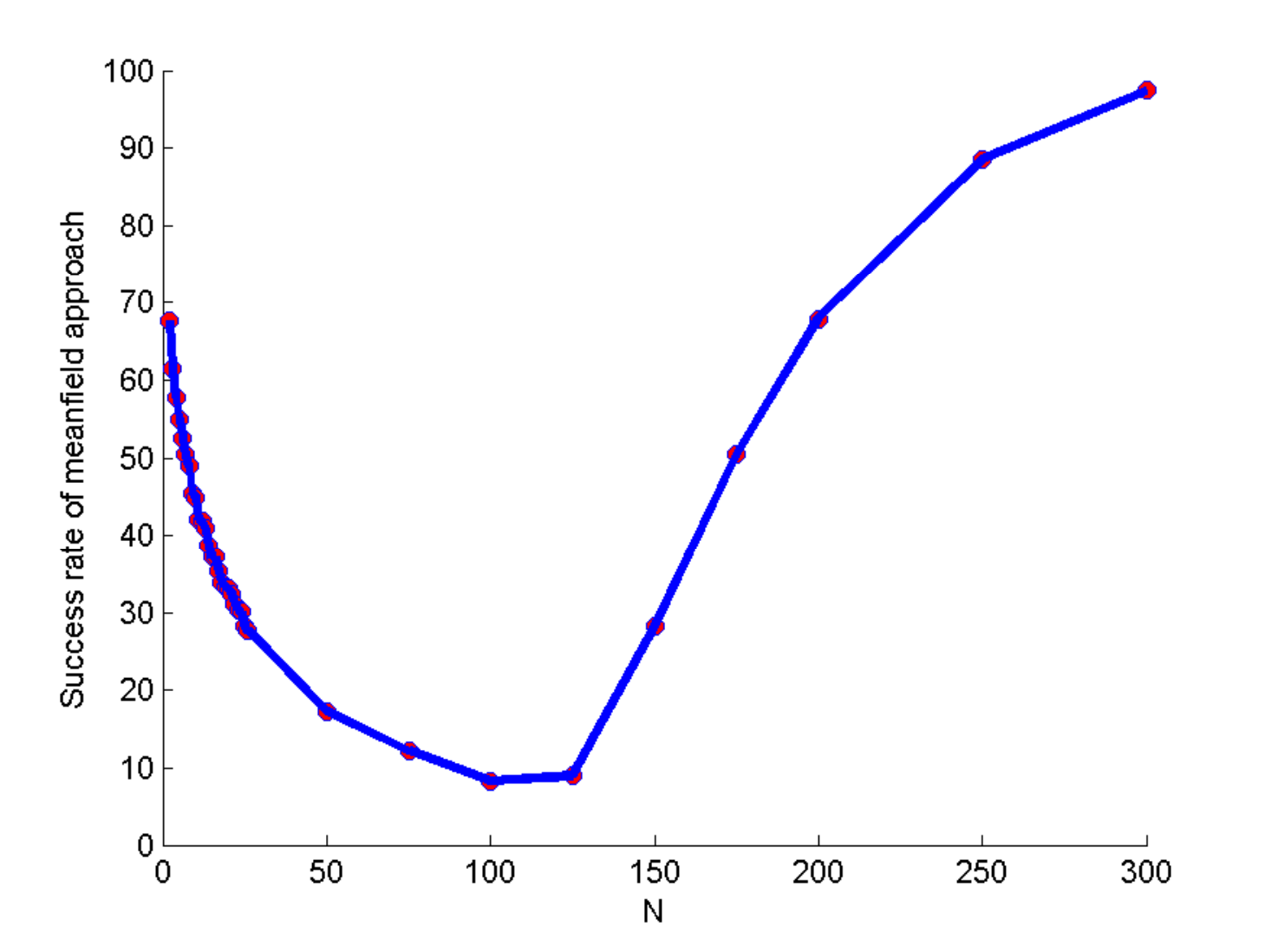}
\caption{Mean field success rate vs. number of spins (N)}
\label{fig:2032}
\end{figure}

\begin{figure} 
\centering
\includegraphics[scale=.5]{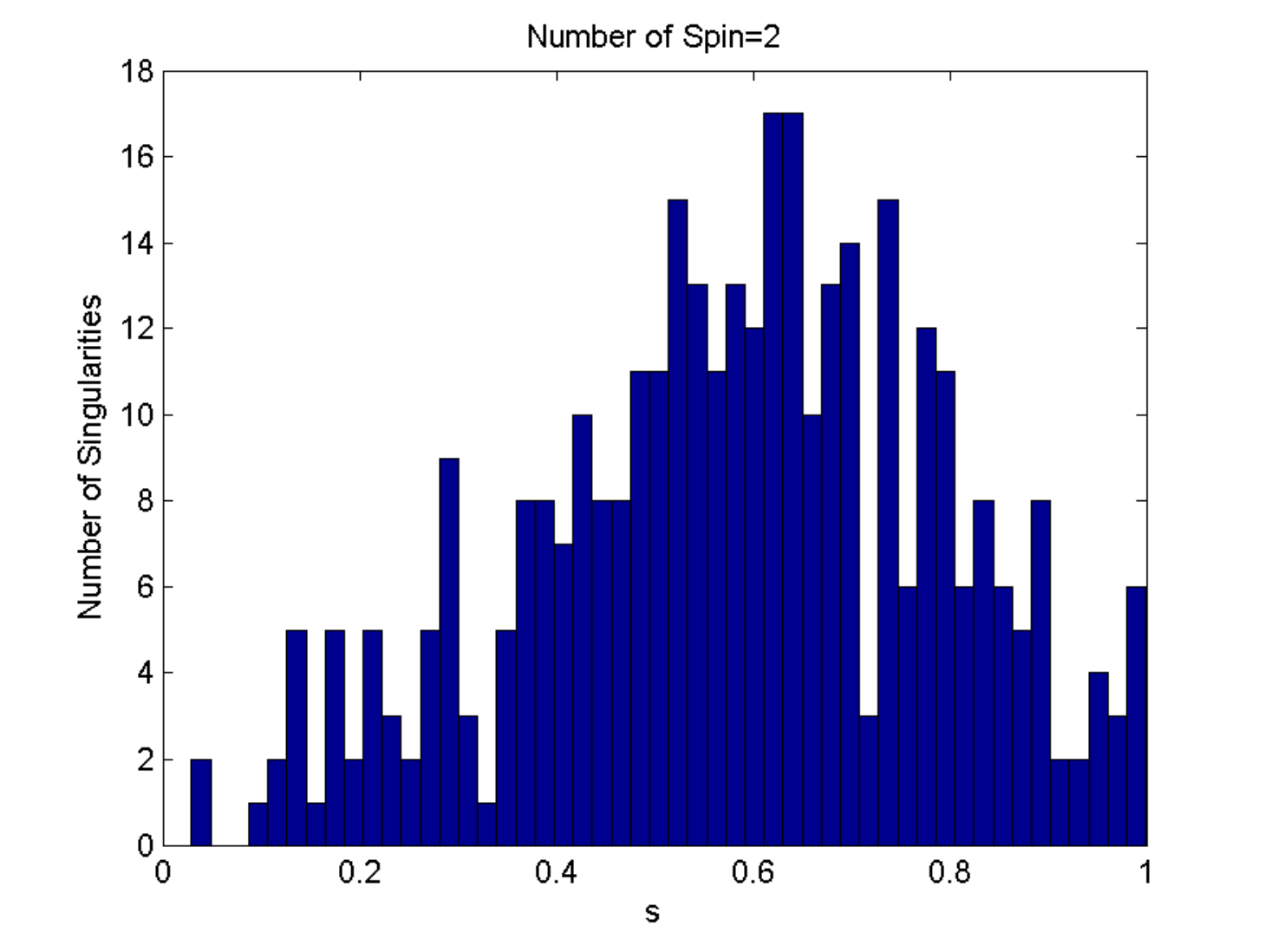}
\includegraphics[scale=.5]{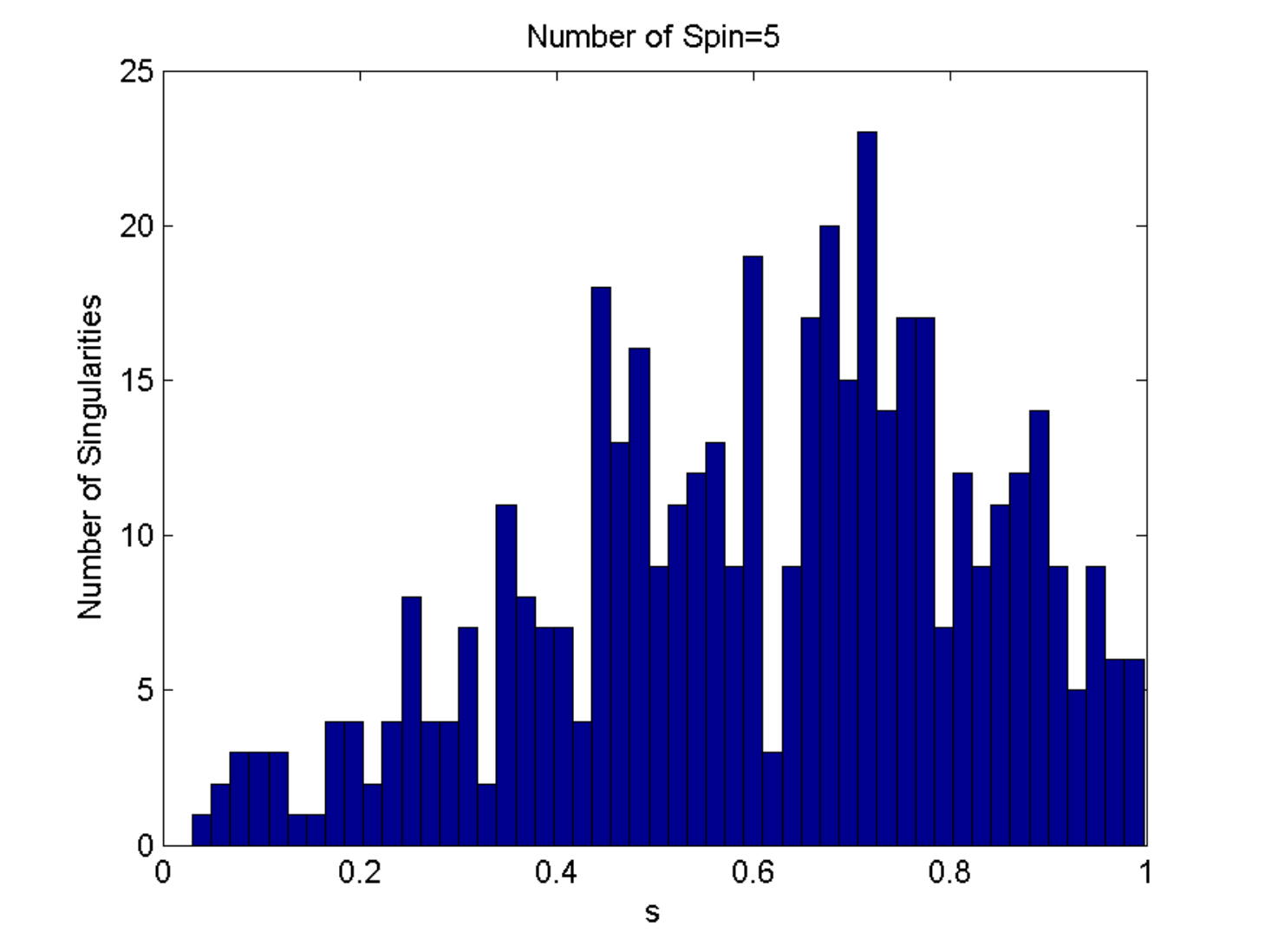}
\includegraphics[scale=.5]{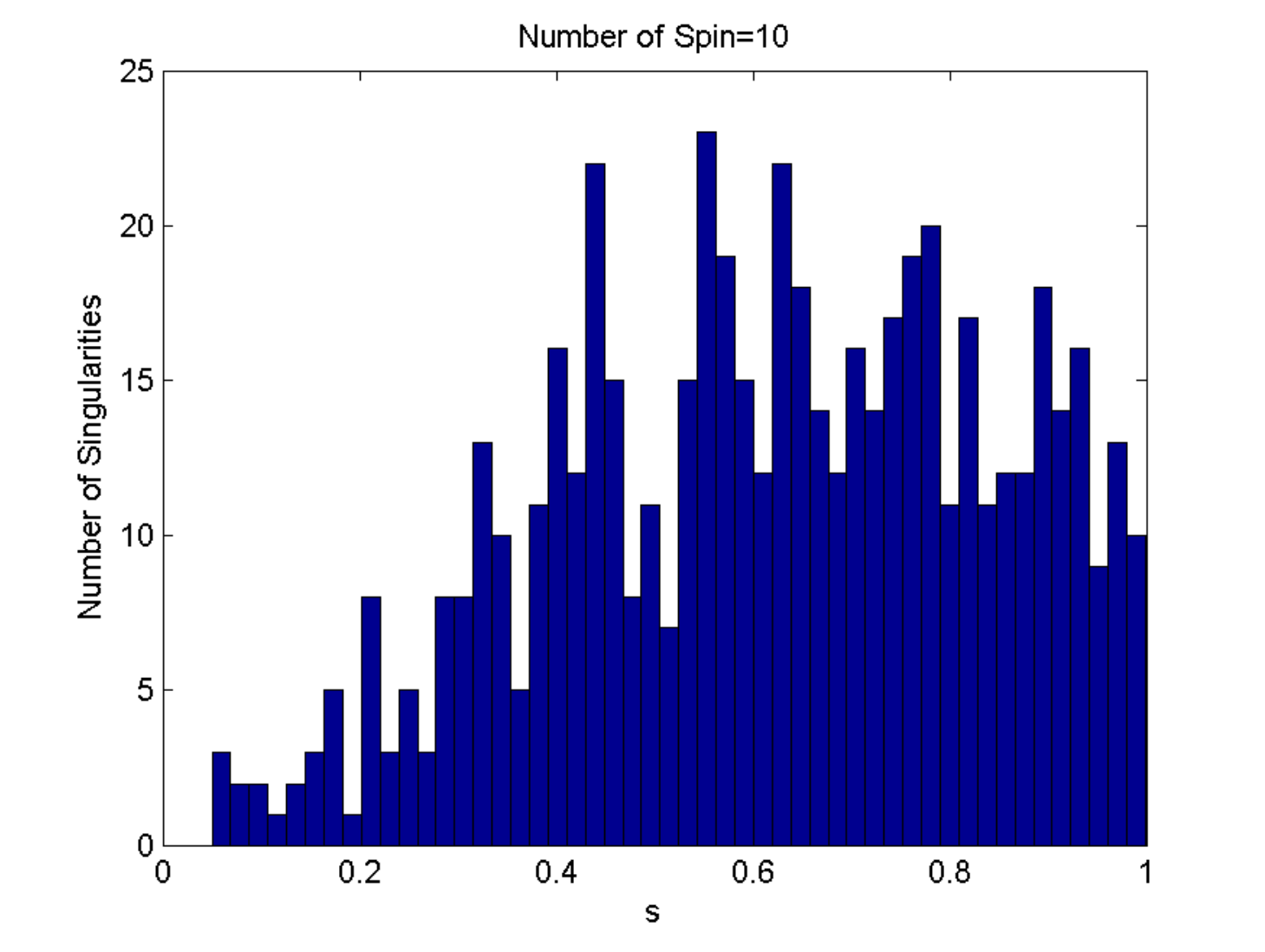}
\includegraphics[scale=.5]{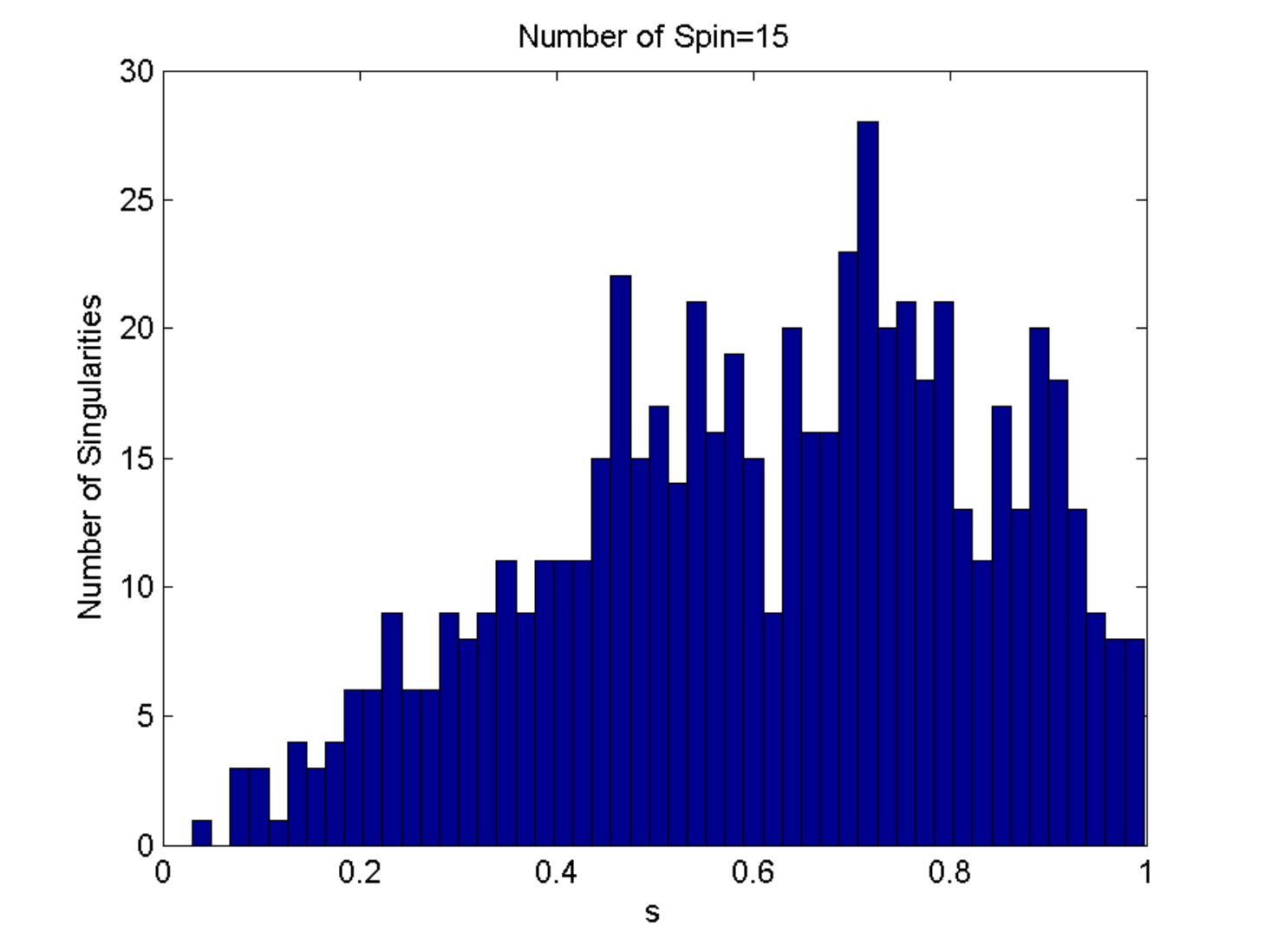} 
\includegraphics[scale=.5]{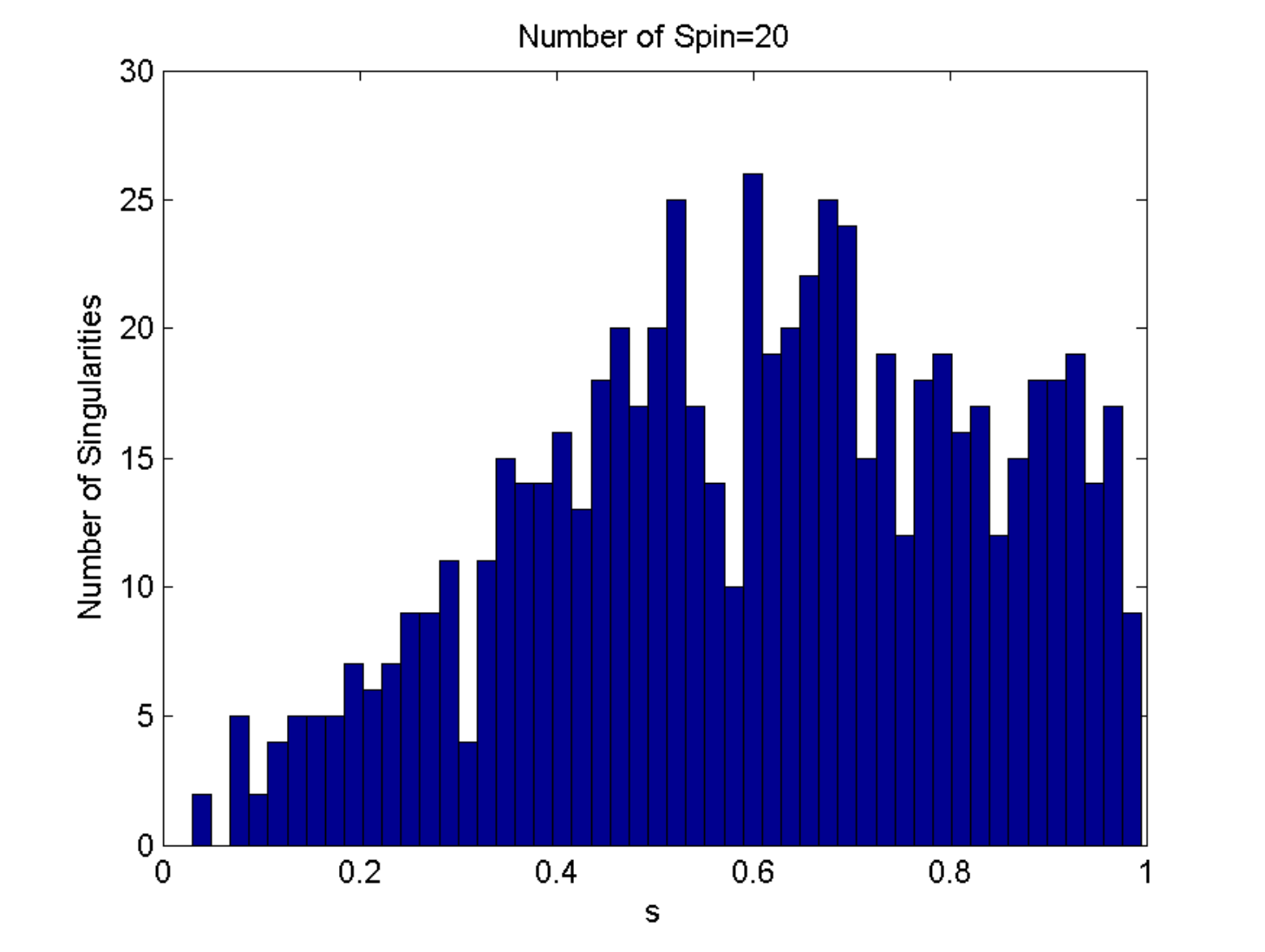}
\includegraphics[scale=.5]{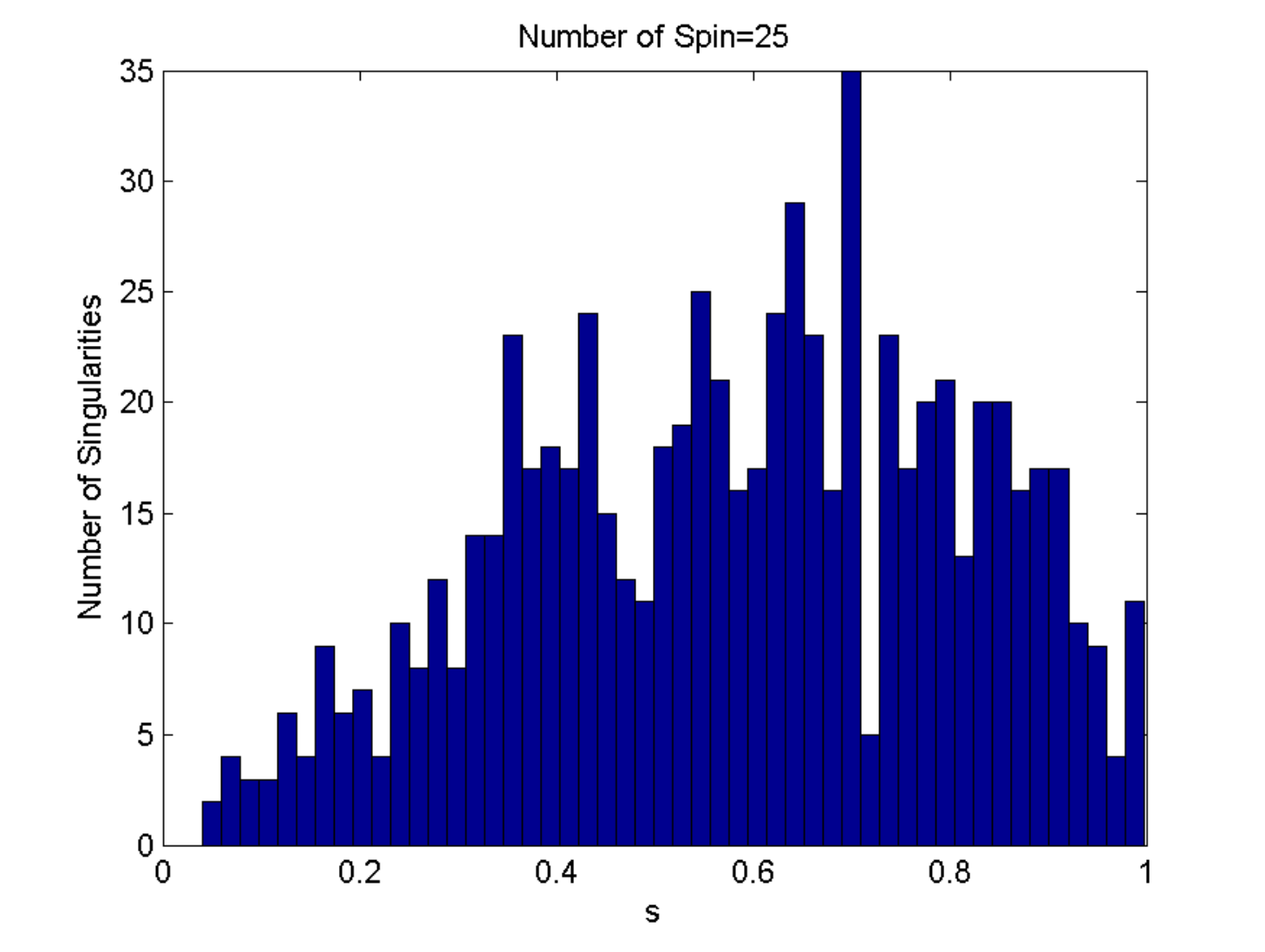}
\includegraphics[scale=.5]{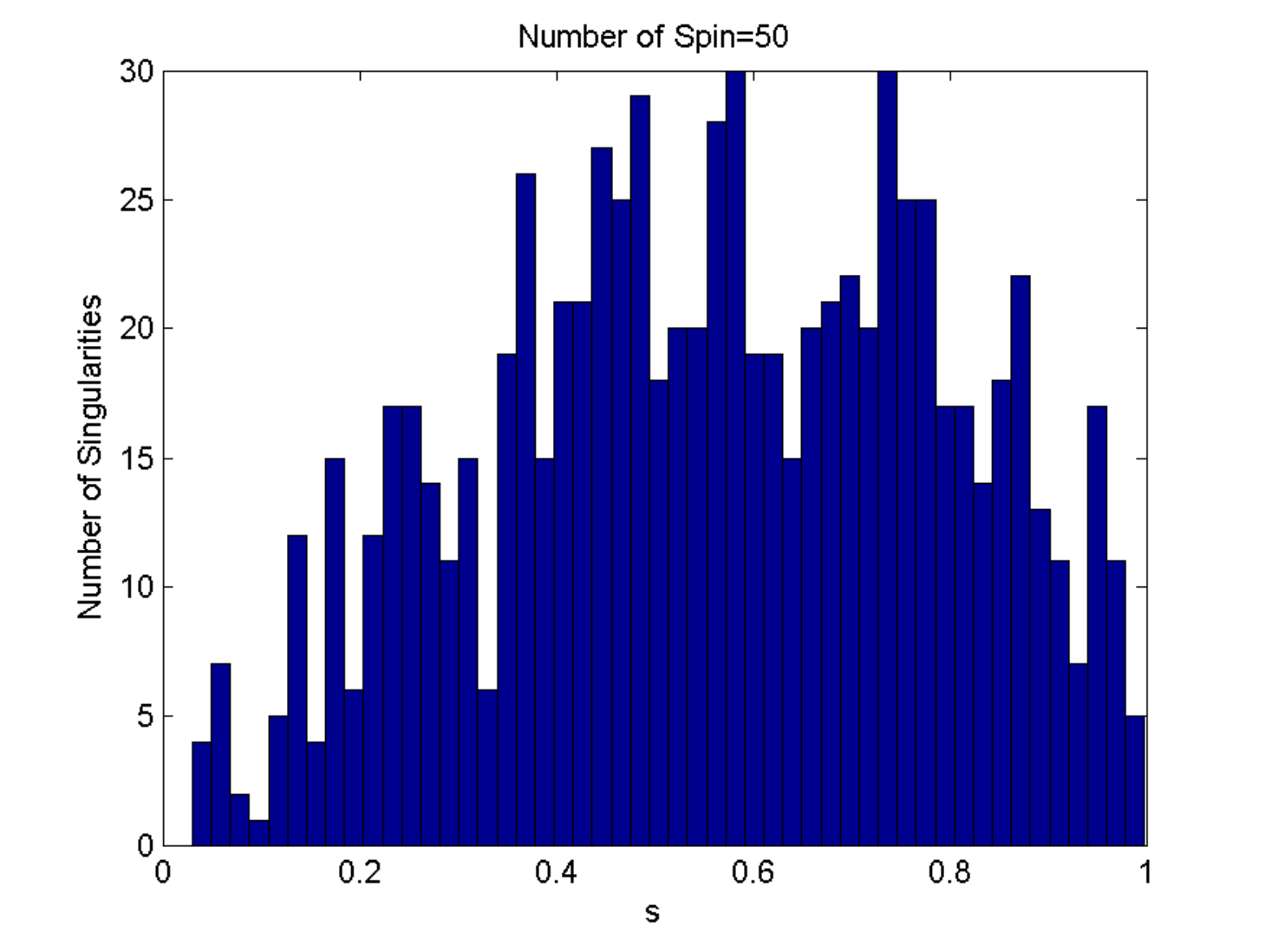}
\caption{Number of singularities in different values of $s$ for small $N$ values.}
\label{fig:204}
\end{figure}

\begin{figure} 
\centering
\includegraphics[scale=.5]{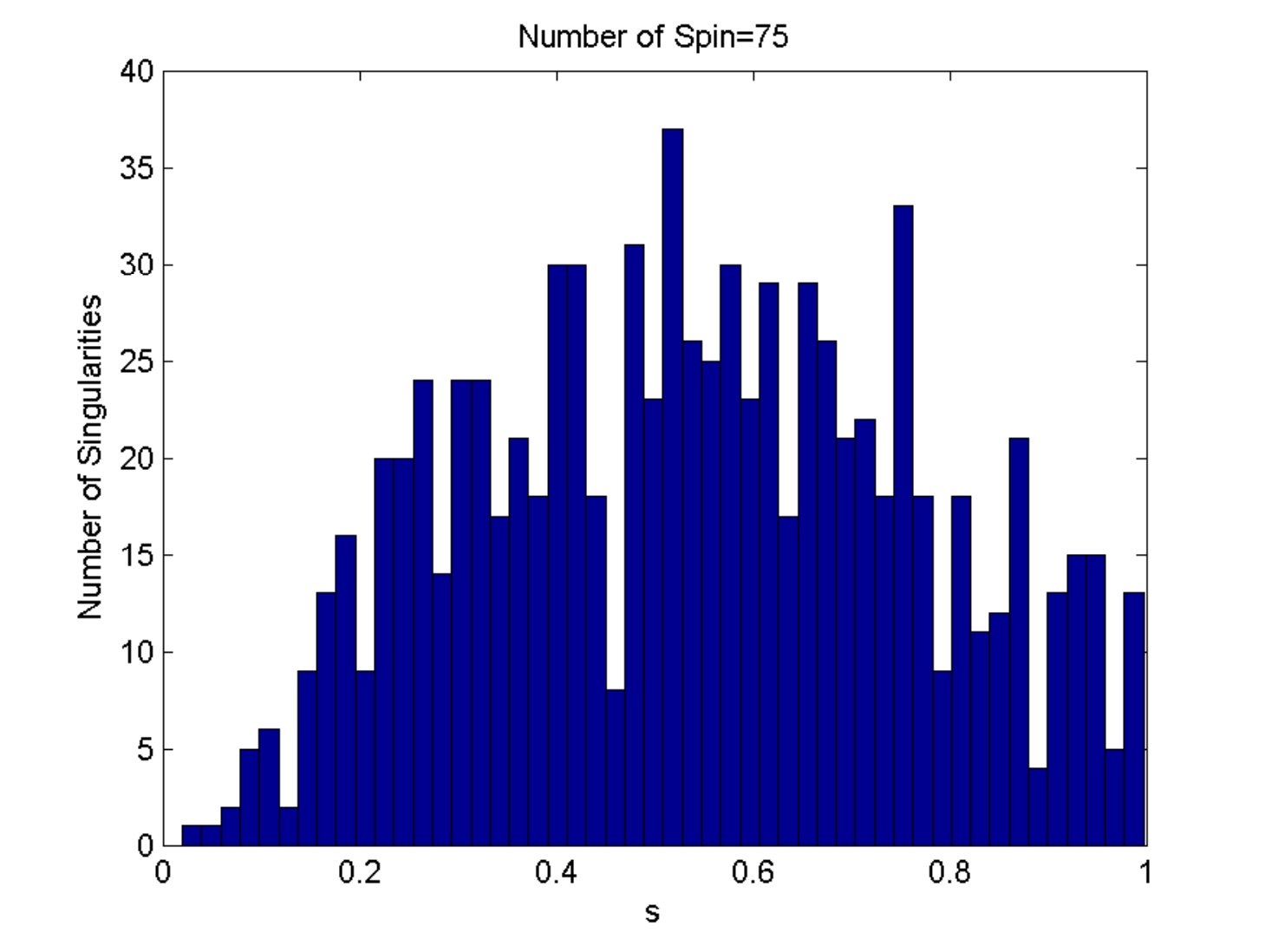}
\includegraphics[scale=.5]{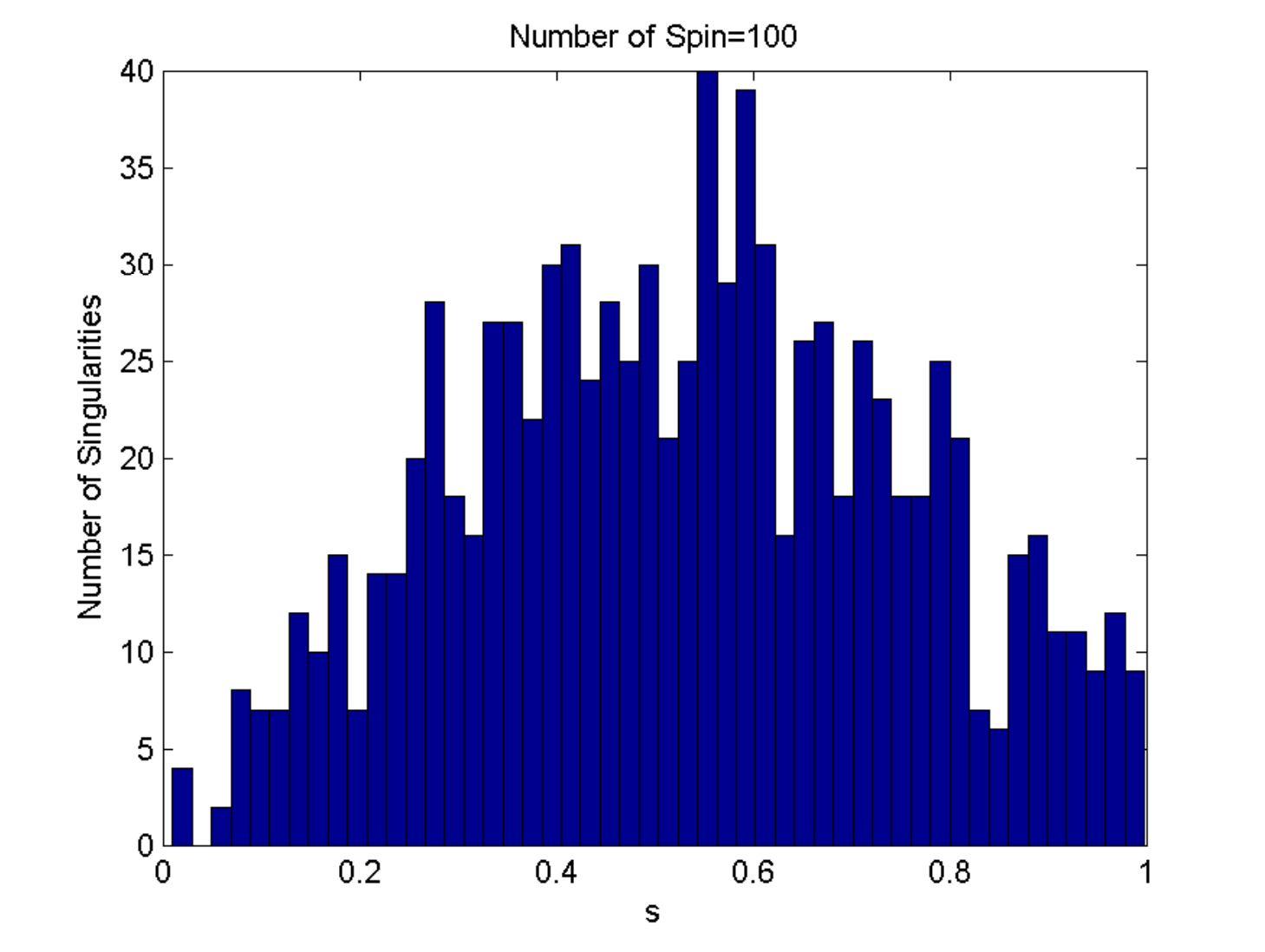}
\includegraphics[scale=.5]{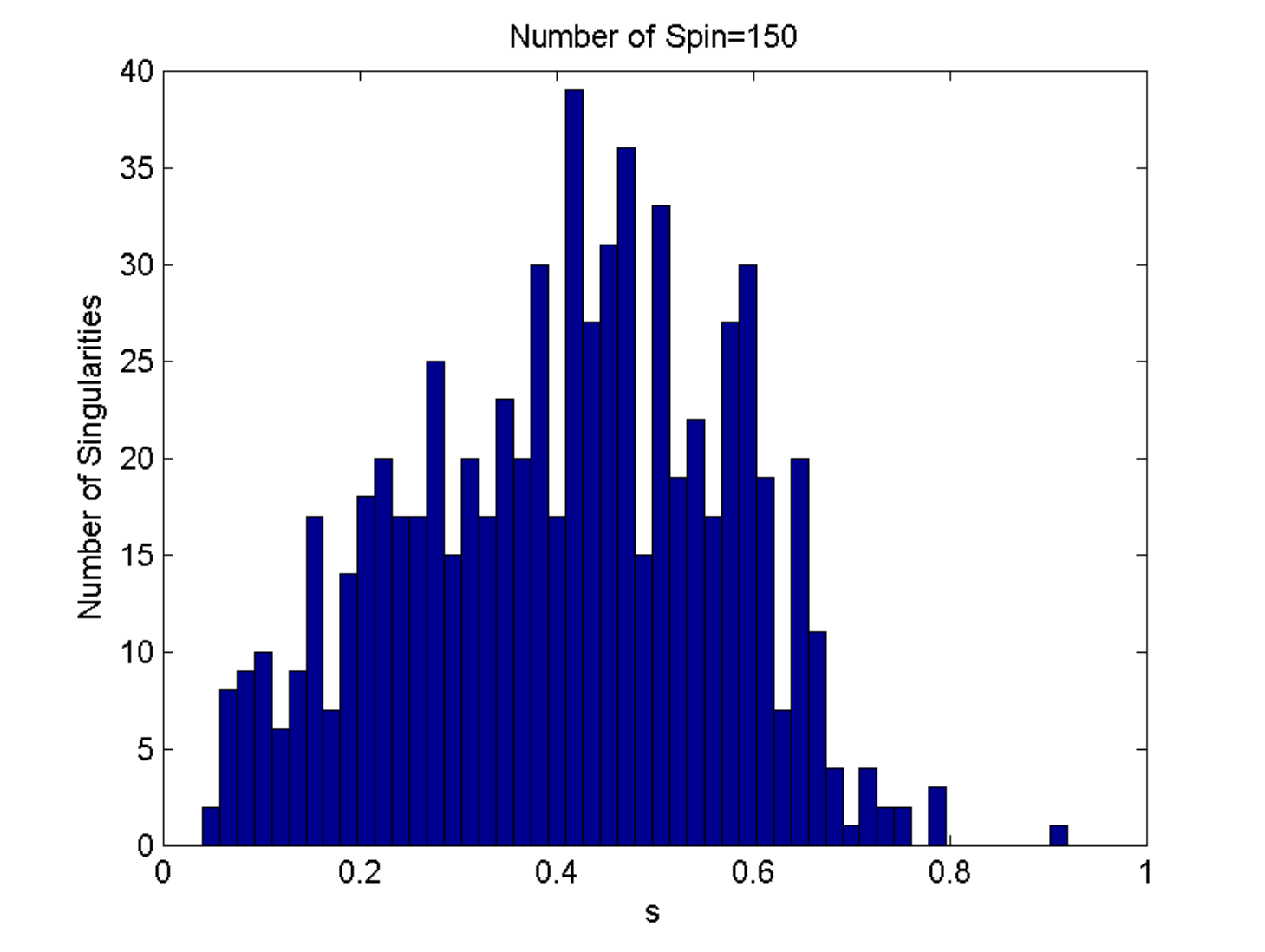}
\includegraphics[scale=.5]{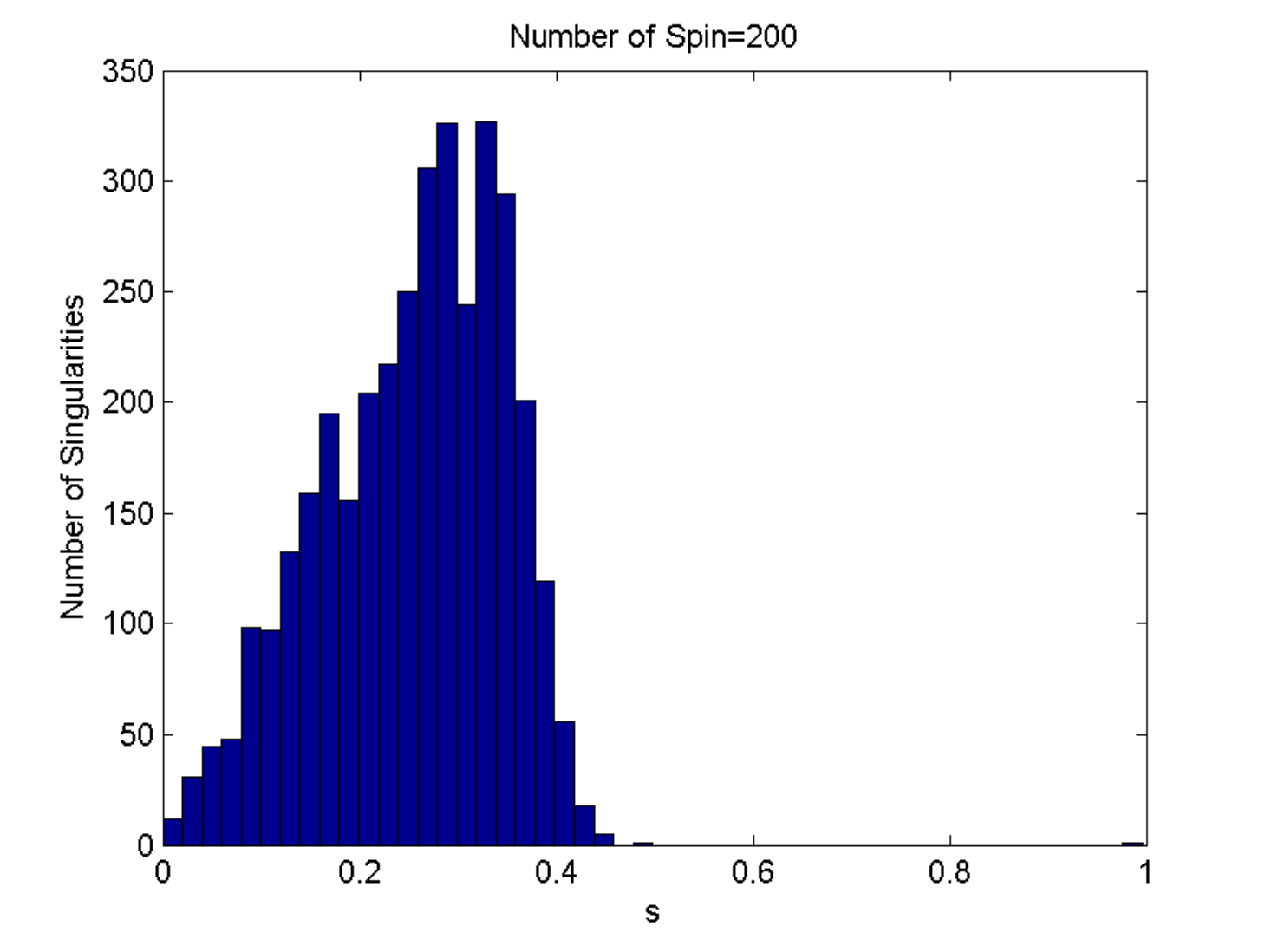}
\includegraphics[scale=.5]{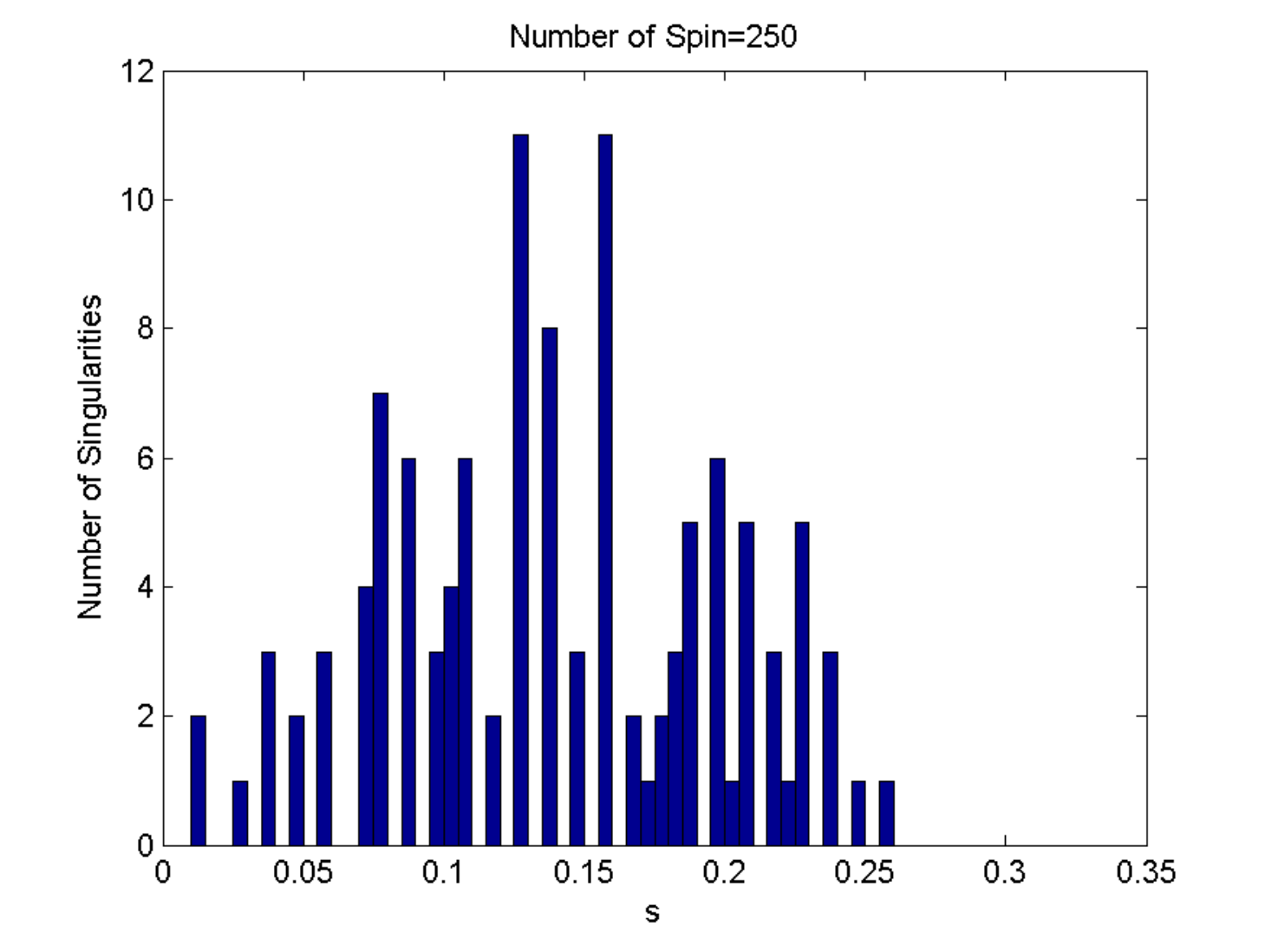}
\includegraphics[scale=.5]{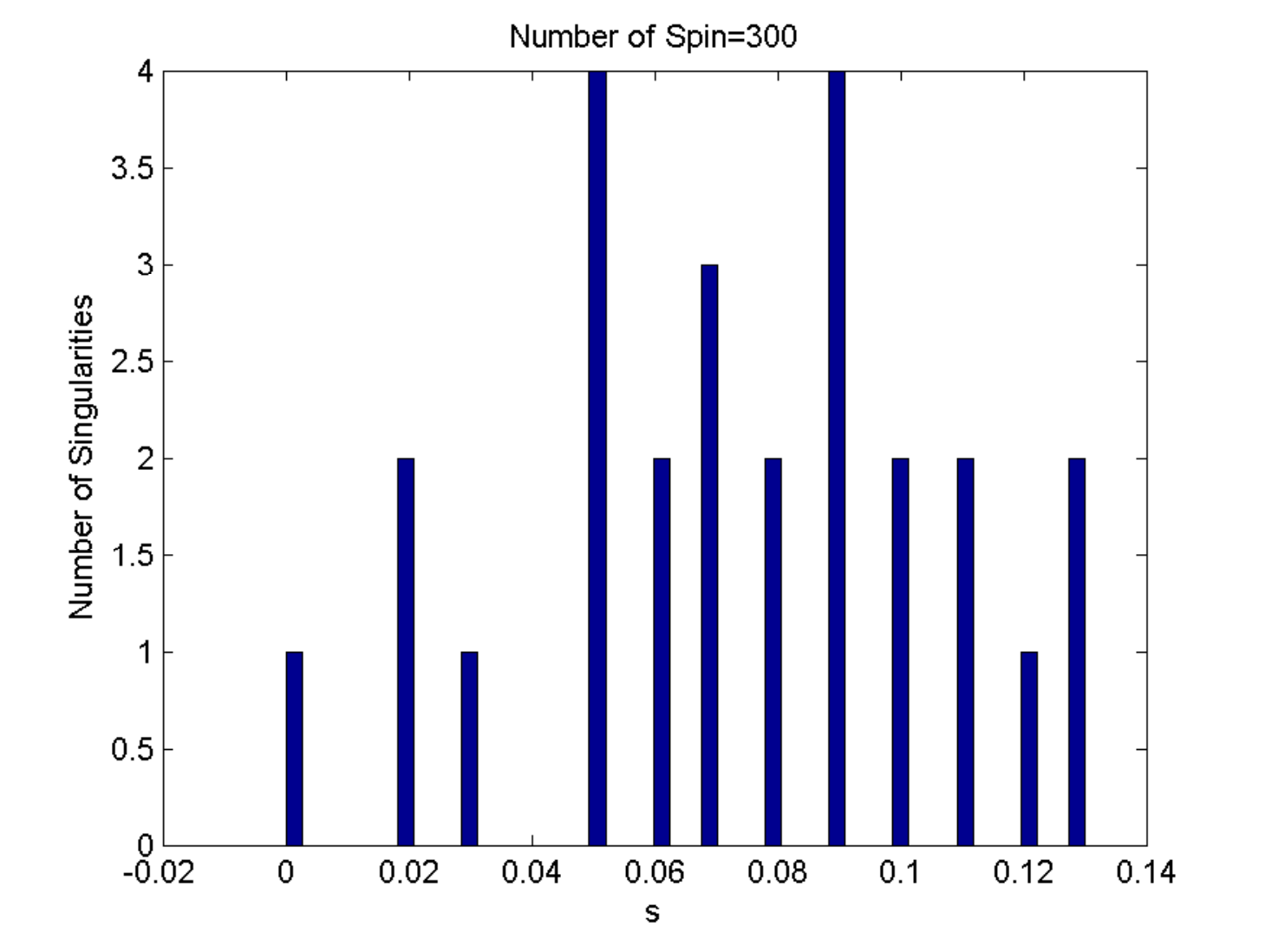}
\caption{Number of singularities in different values of $s$ for large $N$ values.}
\label{fig:205}
\end{figure}

The generalization to a large number of spins is straightforward, but due to the large number of coupling parameters only statistical investigation of the system is possible. We sample over $10000$  realizations for the values ${{J}_{\mu}}, {{J}_{\mu \nu}}$ uniformly randomly distributed within the  interval $[-1,1]$, for different site numbers $N$,  and $Z=N-1$. We solve numerically 
Eq.(\ref{eq:sing}). The success or failure of the lowest order approximation is determined by whether a singularity appears.  The success rates over all the relizations  are plotted  in Fig.\ref{fig:2032}. We observe  that for $N=100$ 
the success rate is at a minimum, and it increases again towards almost  100\% for large $N$. The histograms for the distribution of singularities $s$ are represented in Figs.\ref{fig:204},\ref{fig:205} for various numbers of spins in the system.

We interpret the transition from a decrease to an increase of the success rate  as an interplay between two regimes. 
The first regime of small $N$ corresponds to a situation where the interactions controlled by ${J}_{\mu \nu}$ between the spins 
dominate and their randomness frustrates their orientations. This frustration increases with the spin number $N$ and leads 
to an increase of the failure rate. The second regime takes place once the linear contribution due to ${{J}_{\mu}}$ terms  overcomes 
the interaction energy, the $z$-spin component prefers to be polarized along  
${{J}_{\mu}}$ and the success rate increases again until the linear terms impose the spin choice for $N\to\infty$.
The passage to the transition is noticeable in the histograms where the singularity distribution is broad in the 
frustrated regime whereas it becomes scarce and significant only for small $s$ in the polarized regime.

\section{Conclusions and Perspectives}

We have analyzed the performance of the lowest-order mean field approach when describing  
quantum adiabatic evolution. Even though the success rate is not perfect, this approach 
has the merit of simplicity and can be used as a basis for martingale approach. In contrast to an exact algorithm that  requires exponential (in the number of qubits) resources from the beginning, this approach requires only polynomial resources (though with a finite probability of failure), and the requirements increase gradually in the process of computation.   

The efficiency of the algorithm depends on the number of qubits $N$ and has a minimum in a point
that separates   polarized   and frustrated regimes. These results open a new issue and 
pave the way on the use of many-body approximations not only for a deeper understanding of a quantum computer device 
but also for a general martingale approach for the solution of nonpolynomial discrete optimization problem.
Up to now, this study has been realized in the leading order in the  
$1/Z$ expansion method. The inclusion of the next orders may be promising in a future work in order to improve 
the success rate to a value closer to unity.

\section*{Acknowledgments}
This research was supported by the  National Science Center (Poland) Grant No.2016/22/E/ST2/00555 (ED), EPSRC (UK) Grant  EP/M006581/1 (AZ) and in part by the Russian Ministry of Education and Science via the Increase Competitiveness Program of NUST MISiS Grant No. K2-2017-085 (AZ).

\end{document}